\documentclass[letterpaper, 12pt]{article}

\usepackage{mathrsfs}

\usepackage{amsmath}
\usepackage{amssymb}
\usepackage{amsthm}
\usepackage{epigraph}
\usepackage{graphicx} 
\usepackage{url}
\usepackage{fancyhdr}
\usepackage{color}
\usepackage[czech,english]{babel}
\usepackage{indentfirst} 
\usepackage[T1]{fontenc}
\usepackage{authblk}
\usepackage[utf8]{inputenc}

\usepackage[margin=1in]{geometry}
\usepackage{lipsum}

\usepackage{setspace}

\begin{document}
\begin{titlepage}
   \begin{center}
       \vspace*{1cm}

       \textbf{\Large The old problem of the cosmological constant solved?}

       \vspace{0.5cm}
        
       \textbf{Jan Nov\'ak}

      {\itshape jan.novak@johnynewman.com\\
       Deparment of Physics, Faculty of Mechanical Engineering, Czech Technical University in Prague, Prague, 166 07, \\
       Czech republic}\\

       \textbf{Oem Trivedi}

      {\itshape oem.t@ahduni.edu.in, oem.trivedi@vanderbilt.edu\\
       International Centre for Space and Cosmology, Ahmedabad University, Ahmedabad 380009, \\
       India\\
       Department of Physics and Astronomy, Vanderbilt University, Nashville, TN, 37235,\\
       USA}\\
       28.5.2025
            
   \end{center}

\begin{abstract}
We formulate an approach to quantum gravity, called the ring paradigm. Gravity
 is mediated superluminally, and the graviton is described as a phonon on the grid of
 matter in the Universe. This theory has very interesting applications to cosmology and
 would ultimately solve the old problem of the cosmological constant. It further gives
 new impulses to the scalar field theories because the gravitational ring decays to some
 phantom field. As is obvious, we radically break the Lorentz invariance, which means
 that some generalization of the Haag-Lopuszanski-Sohnius theorem in quantum field
 theory is possible.  
\end{abstract}

\begin{center}
 "Essay written for the Gravity Research Foundation 2025 Awards for Essays on Gravitation"
\end{center}

\end{titlepage}

\selectlanguage{english}


We know today that only 5\% of the energy density in the Universe creates so-called baryonic matter. The universe is composed of 27\% by dark matter and approximately 68\% by dark energy, \cite{Max}. We are not completely surprised that the overall composition of matter is unknown, because we are actually studying what we ourselves are made of, and already quite a few candidates have been suggested for the resolution of this problem. Let us mention, for example, supersymmetric particles, \cite{Wess}, axions, \cite{Weinberg, Wilczek}, mirror particles, \cite{Okun}, and many others. The problematics of dark energy look different. As was stressed by Roger Penrose, the name is already a little bit misleading: it is not dark and it is not technically energy. Further, as we see from recent results, the accelerated expansion of the universe can actually be a quantum gravity phenomenon, \cite{Minic, Perez, DSudarsky, Sorkin, Rafael, Oriti, InnesB, Trout, Singh, Li, Basilakos, Kofinas, Verlinde, JNa1}. This means that the formulation of a model of dark energy could be connected with so-far-unknown features of gravity, which makes this area of research exciting.

If we look to history, after the advent of Newtonian physics, \cite{Newton}, approximately two centuries of mathematical reformulation followed. For instance, let's mention d'Alembert's equations, \cite{Alembert}, Lagrange equations, \cite{Lagrange}, Hamilton equations, \cite{Hamilton}, and others. Finally, we grasped the correct postulates of a new lawyer of physics at the beginning of the 20th century. Quantum theory was reached at in 1900 by Max Planck, \cite{Planck}, and the special theory of relativity in 1905 by Albert Einstein. \cite{Einstein}.

In the 20th century, we can carefully say we witnessed something very similar. Of the remarkable more  mathematic research directions, let us mention, for example, the spinor and twistor calculus, \cite{Peskin, Rindler1, Rindler2, Penrose}, supersymmetry, \cite{Wess}, topological quantum field theory, \cite{Atiyah, Gukov, Schwarz, Witten}, and also non-commutative geometry, \cite{Connes}. One could argue that it was the discovery of string theory by Gabriel Veneziano in 1968 that mainly stimulated these advances, \cite{Veneziano}. But to be honest, a key phenomenological step was missing. Even the revelation of the possible existence of higher dimensions, \cite{Randall}, is not comparable with the revolution at the beginning of the 20th century leading to the discoveries of quantum and relativity theory, \cite{Planck, Einstein, EinsteinA}!  

The urgent need for an implementation of a new phenomenology comes from astrophysics and cosmology. We want to primarily stress the significance of the result of Stephen Hawking in the year 1974, when he published his essay "Black holes are not black", \cite{Hawking}. This work describes that black holes emit thermal radiation and goes to show that even the all-absorbing object - the black hole - does not have an infinite duration. Further, we have started to investigate what happens with all the information, which is sucked out by the black hole. This leads to the formulation of the black hole information paradox, \cite{Preskill, Mathur, Calmet1, Calmet2, Calmet3}. There are the following basic alternatives for how to deal with it: information is simply lost or there exists some remnant after the evaporation of the black hole; we are not interested in these two possibilities because they would not give us potential to formulate some dazzlingly new physics. But the other variant examined is the existence of superluminal signalling, \cite{Polchinski}, and its confirmation would ultimately mean that all our approaches to quantum gravity (QG) must be somehow modified.      

The second result is much celebrated observation of accelerated expansion of the universe in the years 1998 and 1999, \cite{Riess, Perlmutter}, which has also begun to change our view on the development of the whole universe. It is starting to lead us to questions such as, what will really be the fate of the universe trillions of years in the future? A picture begins immediately to unfold in our minds, that even black holes will finally disappear in an ever-faster-expanding universe. And what will actually remain here? Will the evolution of the universe transform into the new aeon according to the conformal cyclic cosmology of Roger Penrose, \cite{CCC}?

The cyclic model of the universe gives us a different point of view, \cite{Steinhardt}. According to it, the universe undergoes cycles of about a trillion years when it expands after the Big Bang and collapses again to the Big Crunch. Probably, the biggest criticism against this model is the existence of the singularity in the Big Bang, whose description within this theory is not entirely credible. But we must remember that this is a classical cosmological model, where we do not use the results of QG, which may change this situation. It has also begun to occur models that build the cosmological inflation inside the cyclic cosmology, \cite{Linde, JNa2}. This is also very interesting because such theories would carry all the positives of the cosmological inflation models together with an explanation of the problem of dark energy. 

But what is actually the most striking fact about these theories would be the existence of phantom fields, \cite{Sami}. We might dismiss that models based on them usually lead to the singularities of the type of a Big Rip, \cite{Prado, Oem}, because it is again just on the classical level of physics, and the cyclic model of Paul Steinhardt and Neil Turok is exactly an example of a working theory, so far. But what will be very important for us is that they are usually connected with superluminal signalling, \cite{Oem7, Oem8, Oem9}. And further, the cornerstone of all theories of QG is that they must describe how the null energy conditions in the singularity theorems of Roger Penrose and Stephen Hawking, \cite{Ellis}, are violated. Phantom fields can lead us exactly in this direction, as was possible to see in the following works, \cite{Oem1, Oem2}.\\   

These were exactly the advances in modern science, which have led us to the formulation of an approach to QG, the ring paradigm (RP). As a solution to the big riddle,
 it began to take shape in our minds... We start from the basic symmetries of string theory, \cite{JNa3, JNa4, Witten2}, and we keep the particle sector of the standard model unchanged,   except for the graviton, which is modelled differently.

Let us consider the following toy-model. Imagine the set of Hopf-linked rings in $\mathbb{R}^3$, which will be logitudinally vibrating. All the particles and fields of the standard model are attached to some end-point of the ring, which is defined by the linkage to the other ring. Gravitons, \cite{Dyson}, are then identified with quanta of the longitudinal vibration of this grid of rings, which means that they are phonons. In other words, we are returning to the idea of some form of aether in physics.

The matter in the Universe in this model simply creates a crystal, and we quantize the vibration of it. This would have far-reaching consequences, mainly for quantum information theory, because gravity is mediated superluminally via these gravitational rings, which are by their nature extremely non-local objects! This fact will not be disprovable by any experiment in the near future, since gravity is the weakest interaction. We could only test RP on the level of consistency.\\

As it is clear from the construction, we don't change the form of Einstein field equations in the high-energy sector. What makes a difference is how quickly gravity is mediated on the full non-perturbative level of physics. We claim that it is by some velocity $c_g\gg c$. We can obtain a bound on $c_g$ when we realize that the observable Universe has diameter of approximately $94.10^9$ pc. Because the rings are preparing in Planck time, the velocity of the spreading of gravitational interaction is minimally $10^{70}$m/s. This is simultaneously a new limit to the maximal velocity for the spreading of information in our Universe!

Now, the new field equations look like the following:

\begin{gather}\label{ring}
\mathscr{R}_{\mu\nu} - \frac{1}{2}\mathscr{R}\mathscr{G}_{\mu\nu} + {\Lambda_r} \mathscr{G}_{\mu\nu} = \frac{8\pi G\mathscr{T}_{\mu\nu}}{c_g^4} = \frac{8\pi G}{c_g^4} (T^m_{\mu\nu} + \mathscr{T}^r_{\mu\nu}),
\end{gather}
where $\mathscr{G}_{\mu\nu}$ is the metric and also all the other quantities have an analogous meaning as in general theory of relativity. There are included both the energy-momentum tensor for the ordinary matter $T^m_{\mu\nu}$, as well as the energy-momentum tensor for the gravitational ring, $\mathscr{T}^r_{\mu\nu}$. The cosmological constant $\Lambda_r$ could be computed from quantum field theory. We neglect the term corresponding for $T_{\mu\nu}$ with respect to the second term on the RHS, so

\begin{gather}\label{ring}
\mathscr{R}_{\mu\nu} - \frac{1}{2}\mathscr{R}\mathscr{G}_{\mu\nu} + {\Lambda_r} \mathscr{G}_{\mu\nu} \approx \frac{8\pi G}{c_g^4} \mathscr{T}^r_{\mu\nu}.
\end{gather}

The picture seems to be correct because there were only gravitational rings at the beginning of the cosmological inflation. These equations transformed soon into the classical equations of general relativity, because $\mathscr{G}_{\mu\nu}$ is unstable and decays to some $g_{\mu\nu}$ Planck time later. On the quantum field theory side, the gravitational ring is quickly decaying to the phantom field:

\begin{gather}\label{phantom}
{R}_{\mu\nu} - \frac{1}{2}{R}{g}_{\mu\nu} = \frac{8\pi G}{c^4} T_{\mu\nu} = \frac{8\pi G}{c^4}(T_{\mu\nu}^m + T_{\mu\nu}^{pf}),
\end{gather}
where $T_{\mu\nu}$ is composed from energy momentum tensor of casual matter $T_{\mu\nu}^m$ and energy momentum tensor of phantom field $T_{\mu\nu}^{pf} = \partial_{\mu}\phi \partial_{\nu}\phi - g_{\mu\nu} \Big(\frac{1}{2} \partial^{\sigma}\phi\partial_{\sigma}\phi + V(\phi)\Big)$. \\

Our approach to QG is, in some sense, very conservative as we don't change the nature of some quantum field. The only difference will be that we include into the concept of quantum field theory some field, which mediates gravity superluminally. Of course, this would have immediate consequences for the group of symmetries of this modified quantum field theory.

We would now need to provide the identification of the theory with field equations (\ref{ring}) with the theory with the field equations (\ref{phantom}). For this it is necessary to equate the impulse of the phonon with the impulse of graviton obtained from the modified theory of gravity with the field equations (\ref{phantom}). We do it by the modeling of gravitons as in string theory. The next step will be to equate the impulse of the graviton in this theory with the impulse of graviton in the modified theory with Einstein field equations (\ref{ring}). We want to note that we could get an approximate result, what are actually gravitons, from gravitational wave physics when we do the splitting of the metric $\mathscr{G}_{\mu\nu} = \mathscr{\eta}_{\mu\nu} + \mathscr{H}_{\mu\nu}$, where $\mathscr{H}_{\mu\nu}$ is a small perturbation and $\mathscr{\eta}_{\mu\nu}$ is a true Minkowski tensor. Then we obtain the impulse of the gravitational wave, \cite{Maggiore}. 

If RP should be a viable model of QG, it must lead to a perturbatively renormalizable theory. Therefore, we must study the interaction of graviton-phonons with other particles. This again leads us to consider the crystal made of rings. A macroscopic grid with very large masses (super-massive black holes) must be created from it, so that the effect of the graviton-phonons on the elementary particle will be observable.

There were considered electrons interacting with phonons as an example in the paragraph 6.10 in \cite{Feynman}. We adopt this model and firstly we will develop the ground state for a fermion system. Then we introduce the concept of holes $b$ with the defining relation

\begin{gather}
b_{\alpha} = a_{\alpha}^+
\end{gather}
 and apply the creation and annihilation operator formalism to electrons placed on a vibrating regular grid of nuclei. The distinction to the model in \cite{Feynman} is that here the crystal does not approximate in a limit a continuum. However, it follows from a careful analysis of these concrete computations that modeling the interactions of graviton-phonons with the other particles will be doable. We obtain no infinities in contrast to "quantizing" pure general theory of relativity. It looks strange that graviton-phonons could have no interaction, for example, with massless particles like photons. But there is an interaction on the fundamental non-perturbative level, when gravitational rings "create a link" with all particles.\\

As we already mentioned, the gravitational ring has two substantial roles. It creates a trajectory for all particles and fields, which means that we need to embed the crystal into the space $\mathbb{R}^3$. The second task is that it mediates gravity via these spring-like objects. Therefore, we obtain a self-evident explanation for the problem of dark energy, because we have some extra energy - mass - in our model, which will be pinpointed with the missing energy (about $68\ \%$).

Let's imagine two galaxies connected by a ring. It is made from some material, which we identify in the classical limit with the emergence of some sort of scalar field. The important point is that we deal with a quantum model when the bunch of rings are created again and again every Planck time later and we observe effectively a spring between the objects. Actually, the detailed model concerns condensed matter physics, \cite{Padmanabhan}. As we describe the "gravitational" material by the springs, the "yield point" corresponds to some moment in the history of the evolution of the Universe, when the galaxies were remoted from each other by some given average distance.

But what is hard to believe, it is possible to find a solution to the old problem of the cosmological constant through the application of RP. We already know that general relativity is a plausible theory up to very high energies, and therefore, shortly before the onset of inflation, we can approximate the correct field equations by (\ref{ring}). The cosmological constant $\Lambda_r$ has exactly the value, that we obtain in high energy physics, and it is roughly $10^{120}$ orders bigger than the effective cosmological constant term, $\Lambda$, which appeared approximately 8 billion years after the Big Bang due to the QG phenomenon via the phantom field:

\begin{gather}
{R}_{\mu\nu} - \frac{1}{2}{R}{g}_{\mu\nu} + \Lambda g _{\mu\nu}= \frac{8\pi G}{c^4} T^m_{\mu\nu}
\end{gather}

It is not surprising that the value of the effective cosmological constant from the cosmological inflation, $\Lambda_r$, possesses many orders of magnitude difference from the value of the constant responsible for the late-time cosmic acceleration, because they have a different origin. The constant $\Lambda_r$ should be connected with the energy in the false vacuum represented by the pure crystal made of rings, which is well-known from quantum field theory. The other, $\Lambda$, is a manifestation of the limit of the firmness of the dark energy material.\\

We can now begin to conclude with some general remarks on the work we have pursued here. RP is an example of Lorentz-breaking theory in the gravity sector, and so we must be very careful for two basic reasons: the Lorentz invariance is one of the fundamental principles that we use in modern physics, \cite{Kostelecky}; it is supported by all the experiments that have been performed so far, \cite{Horava}. There is no evidence that such symmetry has to be broken at high energies. And further, the violation of the Lorentz invariance could have a significant effect on the low-energy physics through interactions between gravity and matter, \cite{Sudarsky}. 

Both of these objections can be answered inside RP, \cite{JNa1, JNa3, JNa4}. First of all, the stringent limits that we have for the possible breaking of the Lorentz invariance are basically bounds on the velocity of graviton-phonons. But these "mediators" travel with the velocity $c$, so there is no discrepancy with the result of our measurements. We need to remember the details of our construction of gravitational rings for the analysis of the second point. Our theory, RP, is built in the space $\mathbb{R}^3$, where the rings are created by the velocity $c_g$. The ordinary particles move on these objects with the maximal velocity $c$, however, we fully use the design of the "Lorentz invariant" theory with the limiting velocity $c_g$. So, for the velocities $v$, $c< v < c_g$, have the transformations, under which is RP invariant, a similar shape as the standard Lorentz transformations with just replacing $c$ with $c_g$, \cite{JNa1, JNa3, JNa4}, and the low energy sector of physics will be not affected.  

Only a little attention has been paid to this type of theories. The reason can be evidently sought in relying on the validity of the "no-go" theorems, concretely the Coleman-Mandula theorem, \cite{Coleman}. However, we use in its proof the mathematical tools that are peculiar to the flat space, and therefore the conclusions could not be generalized to other situations, \cite{JNa4}. And further, we already know that some generalizations of the Haag-{\L}opusza\'nski-Sohnious theorem, \cite{Haag}, which is the "let's go" theorem for the case of the existence of supersymmetry in Nature, are already available. An example could be the usage of more general algebras in this theorem, the color Lie superalgebras or $\epsilon-$ superalgebras, \cite{Coleman, Haag, Traubenberg, Aizawa, Stursberg}. \\

What we simply propose in this essay is a shift in the development of physical theories. The first suggestion was that all the interactions, including gravity, are mediated by the infinite velocity, \cite{Newton}. Then we discovered there is some limiting velocity, by which all the particles of the standard model are travelling, $c$, \cite{Einstein, EinsteinA}. And it was supposed that the same holds for the hypothetical graviton. But when we studied these arguments more carefully, we found out it was only an educated guess that gravity is mediated by the velocity of light $c$. There were no real reasons to suppose the validity of this postulate. So, we corrected it, and we finally came up with the assertion that there is really a limiting velocity $c_g$, by which the information spreads in the Universe. Nevertheless, it is much higher than the velocity of light, $c_g >> c$, \cite{JNa1, JNa3, JNa4}. And this restriction will probably remain with us well into the far future. 

There are two immediate outcomes of our theory. The velocity of gravitational waves $c_{gw}$, \cite{Blas}, is exactly equal to the velocity of light $c$. The reason could be seen from the Newtonian limit of RP, where we model graviton as a phonon, \cite{JNa1, JNa3, JNa4}. Next, we use the deep duality between electromagnetic and gravitational interaction, as should be obvious from the description of graviton-phonons as a vibration of the lattice created from the matter in the Universe. The key observation is the similarity between models of photon and phonon, \cite{JNa3, Feynman}. This would also mean further impulses to the ongoing research concerning Weyl and Kerr-Schild double copy, \cite{Zvi, Luna}. So far, a similarity between classical solutions in gauge theory and gravity theory was observed. But now we suspect a much deeper connection even at the level of the full quantum theory of gravity, \cite{JNa3}. 

Will our paradigm have any practical applications? We claim that, if confirmed, it can have far reaching implications in the distant future. Imagine a hypothetical experiment that we built one observatory measuring the position of the supermassive black hole in the center of our Galaxy on Earth and one similar apparatus by some star in the galaxy of Andromeda. If we now deviate the black hole in our Galaxy from its position, we will get this information in galaxy in Andromeda by the velocity $c_g$, which is much bigger than the velocity of light. These technologies could be also used one day to interstellar travel inside and possibly outside our Galaxy, for example in research into the existence of extraterrestrial life.

The RP stimulates further studies in phantom cosmologies, \cite{Oem1,Oem2,Oem3,Oem4,Oem5}, but the main task after the formulation of this theory will now rest on mathematicians. We have indeed arrived at the gate of one of the greatest discoveries in the history of modern theoretical physics? If we would be able to use RP for finding a solution for such hard problems as is the black hole information paradox and mainly the old problem of the cosmological constant, then it makes sense to further continue in this research program. And we will certainly remember the words of one of the most prominent mathematicians of 20th century, Alexander Grothendieck, that the speed of light had been corrupted by the devil. Therefore, we need more angels, who will free us from this curse.

\end{document}